# Gain and loss induced topological insulating phase in a non-Hermitian electrical circuit

Shuo Liu[1,2,†], Shaojie Ma[1,2,†], Cheng Yang[3,†], Lei Zhang[3], Wenlong Gao[1], Yuan Jiang Xiang[1,*], Tie Jun Cui[3,*], Shuang Zhang[2,*]

[1] International Collaborative Laboratory of 2D Materials for Optoelectronic Science & Technology of Ministry of Education, Institute of Microscale Optoelectronics (IMO), Shenzhen University, Shenzhen 518060, China

[2] School of Physics and Astronomy, University of Birmingham, Birmingham B15 2TT, United Kingdom

[3] State Key Laboratory of Millimeter Waves, Southeast University, Nanjing 210096, China

† These authors contributed equally to this work
*Corresponding author

Corresponding authors: S. Zhang: S.Zhang@bham.ac.uk; T. J. Cui: Tjcui@seu.edu.cn; Y. J. Xiang: yjxiang@szu.edu.cn;

**There have been considerable efforts devoted to the study of topological phases in certain non-Hermitian systems that possess real eigenfrequencies in the presence of gain and loss. However, it is challenging to experimentally realize such non-Hermitian topological insulators in either quantum or photonic systems, due to the difficulties in introducing controlled gain and loss. On the other hand, the wide choices of active circuit components provide us with unprecedented convenience and flexibility in engineering non-Hermitian topological insulators in electrical circuits. Here, we report experimental realization of a one-dimensional (1D) non-Hermitian topological circuit which exhibits topologically protected edge state purely induced by gain and loss. We show that by tuning the value of the positive/negative resistors in the circuit, our system can switch between different topological phase regions. The topological edge states and interface states are observed at the circuit edge and at the interface between a trivial and nontrivial circuit, which are manifested by a prominent impedance peak at the mid-gap frequency topologically robust to variations of circuit parameters. Our work opens a new gateway towards actively controllable topological systems.**

## Introduction

Topological physics is concerned with the study of quantized parameters that are invariant under continuous variations of the system. Originated from condensed matter physics, investigations on topological phenomena have been extended to classical systems such as photonics, acoustics and mechanics. In particular, nontrivial topology in classical systems has shown great potential in applications such as lasers [1,2], manipulation of acoustic [3,4], and mechanical waves [5,6]. Recently, electrical circuits are emerging as a powerful platform for exploring topological physics, allowing demonstration of some of the topological phenomena, such as spin Hall-effect [7,8], Haldane model and magnetic dipole [9], topologically protected edge state in 1-D Su–Schrieffer–Heeger (SSH) model [10,11], Weyl state and Fermi arc surface state in three-dimensional(3-D) [12-14], and higher-order topological states in electrical circuits [15-17]. It is also more convenient to add strong nonlinear effect to topological circuits than in the photonic and quantum systems. By replacing the coupling capacitor in a 1-D SSH circuit chain with two back-to-back varactor diodes, self-induced topological transitions as a function of the input intensity have been realized [18]. More interestingly, operational amplifier (OpAmp) has been introduced to circuit as a gain element to realize non-Hermitian topological bandgap that is fully reconfigurable via tuning the value of resistors [19], and to observe a bulk Fermi-arc state (bulk drumhead states) that connects between the exceptional points (exceptional lines) [20].

Non-Hermitian systems have attracted much attention since the introduction of the concept of parity-time (PT) reflection symmetry by Bender in 1996 [21]. A non-Hermitian system may experience a phase transition at an exceptional point (EP) in the wave momentum, where the eigenfrequencies change from real to complex numbers. In such non-Hermitian systems, complex potential profiles representing balanced gain and loss are the critical factors for achieving real eigenvalues. Due to the difficulty of generating gain and loss in quantum systems, considerable efforts have been made to the study of PT symmetry in photonic systems [22-25]. Combination of PT symmetry and topological physics has led to new interesting physics that do not exist in Hermitian systems [26-28]. In particular, it was recently discovered by Takata and Notomi that topological insulating state can be achieved solely by adding static gain and loss to a topologically trivial structure [29]. They demonstrated theoretically a 1-D photonic lattice with topological insulating phase being controlled by the gain and loss profiles, and showed that the topological edge mode could exist at the boundary of a finite lattice, or the interface between a trivial and a nontrivial lattice, due to the particle-hole symmetry [30], or the pseudo-anti-Hermiticity. However, such gain and loss induced topological systems have not been experimentally observed.

In this work, we present the design and experimental realization of a non-Hermitian electrical circuit whose nontrivial topology solely arises from the introduction of gain and loss. The circuit exhibits four different phase regions at different combinations of the gain and loss settings, three of which support a topological edge state in a nontrivial mid gap opened by the balanced positive and negative resistance. For convenient experimental observation of the edge/interface state from the impedance spectra, a global loss offset is introduced to our circuit for shifting the midgap edge mode to purely real eigenvalue. Circuit simulations and experiments are carried out to confirm the existence of gain/loss induced topological edge state localized at the end of a ten-unit-cell circuit chain.

**Circuit diagram and its characterization**

Here we consider a 1-D non-Hermitian resonator array where the topological phases are solely controlled by gain and loss $\pm g_1$ and $\pm g_2$ elaborately arranged in a pair of resonator dimers, as sketched in Fig. 1a. Gain and loss can be conveniently introduced to electrical circuits in many ways, for example, by employing negative and positive resistors, which enables a direct mapping of the theoretical model presented in Fig. 1a to the electrical circuit as illustrated in Fig. 1b. Each resonant circuit is composed of an inductor *L* and a capacitor *C* connected in parallel, working as a parallel *LC* resonant tank. Resistors $R_a$, $R_b$ and negative resistors -$R_a$, -$R_b$ are added to the corresponding resonant circuits as loss and gain modules, respectively. An additional resistor $R_c$ is added to each resonant circuit as a global loss offset to shift the imaginary part of the eigenfrequency of the edge state to zero, which results in a maximum resonant impedance peak at the edge state. All four resonant circuits are grounded at one end, while the other end is connected to their adjacent ones through two coupling capacitors *C*. Fig. 1c shows the print circuit board (PCB) layout of one circuit unit cell. The non-Hermitian topological circuit is thus constructed by a periodic arrangement of the unit cell in Fig. 1b.

One of the key elements for realizing the gain and loss induced non-Hermitian circuit is the negative resistor, which functions as a power source that pumps energy into the circuit. The negative resistor module, also known as the negative impedance converter with current inversion (INIC), consists of an operational amplifier (OpAmp) accompanied by a negative feedback network, as sketched in Fig. 1d. The virtual open and virtual short circuit conditions between the inverting input and non-inverting input pins of the OpAmp force the input current $I_{in}$ (with magnitude of $V_0/R_c$) out of phase with respect to the voltage $V_0$ applied to it. This implies that the current in a negative resistor flows from low electrical potential to high electrical potential, which is the opposite to a positive (normal) resistor. This counterintuitive behavior of negative resistor makes it consume negative energy from, or pumps positive energy into the circuit, with the amount of electrical energy $P=V_0^2/R_c$, which is exactly the same as what being consumed on a positive resistor $R_0$.

For a circuit with time dependent $e^{i\omega t}$, it can be described by the following equation which links the total input current $I_a$ flowing out of node *a* and the voltage $V_b$ across node *a* and *b*,

$$I_a = \sum_b \left( i\omega C_{ab}(\omega) + \sigma_{ab} + \frac{1}{i\omega} W_{ab}(\omega) \right) V_b = J_{ab}(\omega) V_b \qquad (1)$$

and its matter form, $I = (i\omega C + \sigma + 1/i\omega W)V = J(\omega)V$, where $J(\omega)$ is the circuit Laplacian which contains the complete information of a circuit. $C$, $W$ and $\sigma$ are the Laplacian matrices of capacitance, inverse inductance, and conductance, respectively. The diagonal and off-diagonal components represent the self admittance of a certain node and the mutual admittance between two nodes, respectively. For our circuit, $J(\omega)$ is a complex matrix as it involves resistive terms. The bulk circuit Laplacian $J(\omega,q)$ in the momentum space can be written into a 4×4 matrix by considering the unit cell circuit in the periodic boundary condition (Fig. 1b),

$$J(\omega,q) = i\omega C + \sigma + \frac{1}{i\omega}W = i\omega \begin{bmatrix} 2C+C_0 + \frac{\sigma_1+\sigma_0}{i\omega} - \frac{1}{\omega^2 L_0} & -C & 0 & -Ce^{-iq} \\ -C & 2C+C_0 - \frac{\sigma_2-\sigma_0}{i\omega} - \frac{1}{\omega^2 L_0} & -C & 0 \\ 0 & -C & 2C+C_0 - \frac{\sigma_1-\sigma_0}{i\omega} - \frac{1}{\omega^2 L_0} & -C \\ -Ce^{iq} & 0 & -C & 2C+C_0 + \frac{\sigma_2+\sigma_0}{i\omega} - \frac{1}{\omega^2 L_0} \end{bmatrix} \quad (2)$$

where $q$ is the Bloch wave number linking a certain unit cell with its neighboring unit cells through $V_{n\pm1}(t) = V_n(t) \cdot e^{\pm iq}$. Although the circuit Laplacian in Eq. (2) does not give directly the eigenfrequency of the circuit as itself is frequency dependent, it can be used to analyze the topological invariant and symmetries of topological circuit. The inductive and capacitive components in the diagonal terms are cancelled out at $\omega = \omega_0 = 1/\sqrt{3C_0L_0}$,

$$J(\omega_0,q) = \begin{bmatrix} \sigma_1 & -i\omega_0 C & 0 & -i\omega_0 Ce^{-iq} \\ -i\omega_0 C & -\sigma_2 & -i\omega_0 C & 0 \\ 0 & -i\omega_0 C & -\sigma_1 & -i\omega_0 C \\ -i\omega_0 Ce^{iq} & 0 & -i\omega_0 C & \sigma_2 \end{bmatrix} \quad (3)$$

Without global loss offset $\sigma_0$, the circuit Laplacian $J(\omega_0,q)$ preserves a pseudo-Hermiticity $S^{-1}(q)J(\omega_0,q)S(q) = J^?(\omega_0,q)$, in which $S^{-1}(q) = S(q) = \hat{\sigma}_x \otimes [\cos(k/2)]\hat{I}_2 + \hat{\sigma}_y \otimes [\sin(k/2)]\hat{I}_2$ represents a half-period translation, $\hat{\sigma}_{x,y,z}$ are the Pauli matrices, $\hat{I}_2$ is the 2×2 identity matrix. It also features a pseudo-anti-Hermiticity $J(\omega_0,q) = -\hat{\eta}J^\dagger(\omega_0,q)\hat{\eta}$, where

$\hat{\eta} = \hat{I}_2 \otimes \hat{\sigma}_z = diag(1,-1,1,-1)$ .[31,32] Such a symmetry is known to induce the nontrivial topology forcing pairwise eigenvalues $j(\omega_0, q) = -j^*(\omega_0, q)$. Note that these symmetries are similar to those in the quantum systems [29] due to their similarity in structural configuration.

A new topological invariant, the normalized Berry phase $W = \sum_s \frac{i}{2\pi} \left( \frac{1}{2} \int_{-2\pi}^{2\pi} dq \langle\langle \psi_{B,s} | \psi_{B,s} \rangle \right)$ [29,33], is introduced to determine the topology of the circuit under different gain and loss configurations, which is obtained by firstly calculating the inner product of the right and left eigenvectors in a 4π-period in the momentum space, and then summing over all bands. Here, $|\psi'_{B,s}\rangle$ and $|\psi'_{B,s}\rangle\rangle$ are the right eigenvectors and corresponding left eigenvectors of $J(\omega_0, q)$, and they should be normalized as $\langle\langle \psi'_{B,s} | \partial_q | \psi'_{B,s} \rangle / \langle\langle \psi'_{B,s} | \psi'_{B,s} \rangle$ before sent for the calculation of $W$ [34]. In contrast to the original definition of Berry phase for the 1D model which is calculated within one loop of the first Brillouin zone, here the loop spans two rounds of the first Brillouin zone. It is calculated that $W$ equals 1 for $\sigma_1 \sigma_2 > 0$, indicating a topologically nontrivial state, while turns zero for $\sigma_1 \sigma_2 < 0$, indicating a trivial state.

To obtain the eigenfrequency of the circuit, we further construct the Hamiltonian from the circuit Laplacian matrix $C$, $W$, $\sigma$ (Note S2 in [35] and Eqs. (S7)-(S9) in the new basis of $\psi(t) = (V(t), \dot{V}(t))$,

$$H = i \begin{bmatrix} \frac{\sigma}{C} & \frac{W}{C} \\ -I & 0 \end{bmatrix} \qquad (4)$$

It is proved that the topological invariant of the circuit calculated from both the circuit Laplacian and circuit Hamiltonian are identical [19]. As explained in [35] Note S2, circuit Laplacian and circuit Hamiltonian are connected in a way that the zeros of eigenvalue spectra of circuit Laplacian correspond to the eigenvalues of the circuit Hamiltonian (i.e. eigenfrequencies of the circuit) [13,15,19].

**Bulk properties at different phases**

To analyze how the band structure evolves among different phases at different gain and loss settings ($\sigma_1$, $\sigma_2$) under a fixed global loss offset $\sigma_0$=2.1 mS, we plot in Fig. 2a the phase diagram of the bulk circuit model, which is numerically obtained by analyzing the band structure at 601×601 data points in the $\sigma_1$-$\sigma_2$ plane ranging from 0 to -6 mS. The band structure can be divided into four regions in the $\sigma_1$-$\sigma_2$ plane, labeled as I, II, III, and IV, respectively, based on whether there exist crossings between the upper two bands, and between the lower two bands, and the degeneracy at zero momentum *(q=0)* between the 2$^{nd}$ and the 3$^{rd}$ band. Note that phase I is located on the *x* and *y* axes from 0 to -4mS, as indicated by the light blue line. When all resonators have identical loss, that is, with only a nonzero global loss $\sigma_0$=2.1 mS while $\sigma_1$=0, $\sigma_2$=0, no bandgap is found either in the real part (Fig. 2b) or the imaginary part (see [35] Fig. S1a) of the band structure, with the second and third bands touching at $\omega_0$=620 KHz, forming a Dirac-like dispersion due to the band folding at *q*=0.

Interesting physics emerge as we add distributed gain and loss to the circuit, that is, when $\sigma_1 \cdot \sigma_2 \neq 0$. Here we investigate the band structure of the proposed non-Hermitian circuit for the four points in Fig. 2a located in the four different phase regions. Due to the presence of nonzero global loss $\sigma_0$, all four bands for any points in the four regions remain complex in the entire first Brillouin Zone (see [35] Fig. S2 for the imaginary part). For point A with $\sigma_1$=-3mS, $\sigma_2$=0 located in phase I, the degeneracy between the middle two bands at *q* = 0 (Fig. 2c) persists as long as one of $|\sigma_1|$ and $|\sigma_2|$ is zero and the other one remains below 4 mS. There are two linear crossings between the real parts of the two upper bands ($\psi_3$ and $\psi_4$), and between that of the two lower bands ($\psi_1$ and $\psi_2$). By setting $\sigma_1$=-3mS, $\sigma_2$=-0.2mS, the circuit enters phase II, as manifested by a lifting of the degeneracy between the middle two bands at zero momentum $q=0$ (Fig. 2d). The crossings of the upper two bands and lower two bands still exist at phase II, but they shift towards smaller momentum $q$ due to the increased gain and loss profiles. As we further decrease $\sigma_2$ to -1.5mS, the system enters phase III, identified by the vanishing of the crossings in the real part of lower two bands ($\psi_1$ and $\psi_2$) in the first BZ (Fig. 2e). Finally, by setting $\sigma_1$=-6mS, $\sigma_2$=-3mS, the system enters phase IV, showing four completely seperated bands in the entire BZ (Fig. 2f). Details of the band structures with zero global offset and their mode pattern analysis, as well as more phase regions of the circuit are given in [35] Note S1. Since the circuit Laplacian at the resonant frequency $\omega_0 = 1/\sqrt{3C_0 L_0}$ takes the same form as the Hamiltonian of the quantum system in Ref. 29, the eigenvalue of circuit Laplacian in the momentum space resemble the band structure of the corresponding quantum model, as detailed in [35] Figure S6. The 2.1mS shift observed in the real part of admittance is due to the global loss offset.

**Nontrivial boundary state at chain edge**

The nontrivial topological feature of the 1-D non-Hermitian topological insulator is manifested by its distinct edge state appearing at the boundaries of a finite chain. However, it should be noted that the observables of topological circuits are different from those in the quantum and photonics systems. Topological circuits are commonly studied through a two-point impedance measured between two adjacent nodes $a$ and $b$, subject to an external current excitation $I_0$ flowing through them [11-13], which is expressed as:

$$Z_{ab}(\omega) = \frac{V_a - V_b}{I_0} = \sum_n \frac{|\psi_{n,a} - \psi_{n,b}|^2}{j_n(\omega)} \tag{4}$$

in which, $\psi_{n,i}$ ($i=a$ or $b$) and $j_n(\omega)$ are the eigenstates and eigenvalues of $J(\omega)$, respectively. As the roots of $j_n(\omega)$ correspond to the eigenfrequencies of the circuit, $Z_{ab}(\omega)$ diverges when the denominator $j_n(\omega)$ crosses zero. Hence, every pole in

$Z_{ab}(\omega)$ represents a mode (either bulk or edge mode) in the finite circuit. Hence, an edge state could be identified at the circuit boundary by a strong resonant peak in the impedance spectra at $\omega_0$.

To demonstrate the existence of topological edge state in our non-Hermitian circuit, a circuit chain with 10 unit cells illustrated in Fig. 3a is fabricated using PCB technology (Fig. 3b). A vector network analyzer (Agilent 8753ES) operating from 30KHz to 6GHz is used to measure the impedance spectra between each adjacent resonators. To provide an appropriate frequency range for the OpAmp and to obtain a distinct impedance peak in the measurement, the values of the inductors and the capacitors are chosen as $C$=470 pF, $L$=470 µF. To be consistent with experiment, we have considered in the theoretical calculations a Q-factor of ~40 for the inductors by adding an imaginary part to the inverse inductance matrix $W$.

In the experiment, A high-speed low-distortion voltage feedback amplifier (Taxes Instrument, LM6171) is used to construct the negative resistor module, which generates the same amount of energy ($P=V*I$) of what a positive resistor consumes. The negative resistor is experimentally tested to have a small deviation of the nominal resistance (4%), which falls in the typical tolerance range of a commercial resistor. Wire wound inductors with (Murata, 2% tolerance) with Q-factor of over 40 (at 600KHz) and 4Ω DC resistance were chosen to help achieve high precision circuit response and sharp impedance resonance. A 4Ω DC resistance is set for the inductors. In the measurement, a DC power supply (Agilent E3648A) having two independent channels provides ±15V DC voltage for the OpAmp. 2.2µF and 2 pF capacitors are connected in parallel with the DC supply pin of the OpAmp to minimize ripple current. A vector network analyzer (Agilent 8753ES) working from 30KHz to 6GHz is used to measure the impedance spectra between each adjacent resonators. As the OpAmp is a nonlinear circuit element, its operational status is highly dependent on the input power. In the experiment, the input power is finely adjusted to 8.61dBm to achieve a maximum resonant impedance at the edge state eigenfrequency 612KHz. In the numerical simulation, we employ the realistic PSpice model for the OpAmp (LM6171 PSPICE Model) provided by Texas Instruments.

Initially, we set the system in phase III by letting $\sigma_1$=-3mS and $\sigma_2$=-1.5mS, which corresponds to the resistance of $R_1$=330 Ω and $R_2$=680 Ω in Fig. 1b, respectively. Fig. 3c shows the real part (blue circle) and imaginary part (black diamond) of the calculated eigenfrequencies of the finite circuit chain. A pair of mid-gap eigenmodes appears around 616 KHz, corresponding to two localized edge states at the left and right ends of the circuit chain. The two edge modes are also indicated by two isolated curves in the real part of eigenvalues of the finite circuit Laplacian, as shown in Fig. 3d. Since the eigenvalue of $j_n$ for the left edge mode crosses zero at around 616 KHz (see [35] Fig. S8a), one expects a strong localization of the impedance at the left edge. Fig. 3e and 3f show the measured and simulated spectrum of impedance measured between all adjacent nodes, respectively, and they match reasonably well with each other. The impedance spectrum between the two leftmost nodes (red curve) shows a distinct peak over $10^5$ Ω at 616KHz, while the impedances measured between all other adjacent nodes (black curves) are much lower. The slight discrepancy between the measurement and simulation is mostly due to the non-ideal operating status of the OpAmp in the real circuit network, which is difficult to precisely predict in numerical simulations using PSpice model (see method). To directly observe the localization effect of the nontrivial edge state, we present in Fig. 3g and h the measured and simulated impedance distributions at the edge mode frequency, respectively. The presence of nontrivial edge state is indicated by a prominent impedance peak localized at the left edge.

In order to excite the right edge mode, one needs to change the global loss offset to $\sigma_0$=0.7mS based on our calculation. Fig. 4a shows the calculated eigenfrequencies for the same finite circuit chain with a global loss offset $\sigma_0$=0.7mS. In this case, the imaginary part of the right edge mode is shifted to zero at 618KHz (Fig. 4a), as shown by Fig. 4b, due to a vanishing eigenvalue $j_n$ (see [35] Fig. S8b). Fig. S8c shows the simulated impedance spectra between all adjacent nodes. As expected, we observe an impedance peak of about $10^5$ Ω between the two rightmost nodes at the edge frequency of 618KHz, as highlighted by the red curve, which agrees well with the theoretical results calculated from circuit Laplacian (Fig. 4d). The numerically (frequency domain solver) and theoretically calculated impedance distributions shown in Fig. 4e and f further reveals the tight localization of the impedance at the right edge.

However, for the right edge mode with a smaller $\sigma_0$=0.7mS, the edge mode as well as some of the bulk modes are in the lasing mode, as observed from the sorted eigenfrequencies in Fig. 4a. Such modes with negative imaginary parts are unstable in our circuit because the OpAmp cannot supply continuously increasing power, and thus does not produce the desired circuit responses as shown in Fig. 4 simulated with frequency domain solver. Hence, the right edge mode is considered to be non-physical. While such an issue does not occur in the case of left edge mode as the imaginary part of all modes are negative (Fig. 3c) due to the larger global loss offset ($\sigma_0$=2.1mS).

**Nontrivial interface state between a trivial and nontrivial chain**

Similar to the SSH model in condensed matter physics, the nontrivial edge state exists not only at the edge of a nontrivial lattice, but also at the interface between a nontrivial and a trivial circuit chain, as illustrated in Fig. 5a, where both chains contain five unit cells. The trivial circuit chain is realized by inverting the sign of $\sigma_1$. In the first case, the left chain is set in phase IV with $\sigma_1$=-6mS, $\sigma_2$=-3mS. A pair of mid-gap modes appears at around 608 KHz (Fig. 5b), in which the left and right mode correspond to the edge mode at the left boundary and the interface mode at the center of the chain, respectively. To observe the interface state, we offset the imaginary part of the right edge mode to zero by adding a global loss offset $\sigma_0$=2.1mS. Now the upper isolated eigenvalue curve in Fig. 5d crosses zero at the interface mode frequency 608KHz, which can also be confirmed from the magnitude of $j_n(\omega)$ in Fig. 5f. We observe a distinct impedance peak at the center of circuit which decays exponentially into the bulk (Fig. 5h).

In the second case, the left chain is set in phase III with $\sigma_1$=-3mS, $\sigma_2$=-1.5mS. Similar to the first case, a pair of interface modes emerges between the gap of bulk modes at around 620 KHz (Fig. 5c). In this case, the left one corresponds to the interface mode, which is shifted to purely real eigenfrequency via the global loss offset $\sigma_0$=0.2mS. This is confirmed in Fig. 5e, in which the upper one is shifted to zero admittance at the interface mode frequency, resulting in a vanishing eigenvalue $j_n$ in the magnitude of $j_n(\omega)$ (Fig. 5g). We observe from Fig. 5i that the impedance peak which identifies the interface state exhibits a slower decaying rate than the first case. This is due to the appearance of bulk modes near the interface mode frequency, as can be observed from real part of $j_n(\omega)$ in Fig. 5e. As a consequence, the slope of the localized impedance peak in the impedance distribution plot (Fig. 5i) is not as steep as the first case. Note that these two cases are not physically realizable due to that some of the bulk modes are in the lasing mode.

**Conclusion**

To conclude, we have proposed and experimentally demonstrated a 1-D non-Hermitian topological circuit in electrical circuits, which exhibits highly pronounced impedance resonance at the boundary of the nontrivial lattice. The topological phase is highly robust against local disorder of the circuit components of 5-20% tolerances (See [35] Note S3). Importantly, one can conveniently introduce nonlinear effect to the non-Hermitian topological circuit by employing nonlinear circuit elements, for instance, varactor diodes [18, 36], transistors, and operational amplifiers. Topological circuit are currently providing us a convenient experimental platform for exploring new topological physics in the classical regime. Our work may pave the way for tunable topological circuits by properly utilizing active components such as photoresistor, whose resistance is controlled by the light intensity. The wide range of circuit functions of OpAmp (including addition, subtraction, integration, and differential, etc.) also provides highly flexible configurations for the hopping parameters design, promising interesting designs with unusual topological properties.

**Acknowledgements**

This work was funded by the European Union's Horizon 2020 Research and Innovation Programme under the Marie Sklodowska-Curie Grant Agreement No 833797, the Wolfson Foundation, Horizon 2020 Action Project No. 734578 (D-SPA), the National Key Research and Development Program of China (Grant No. 2017YFA0700201), in part by the National Natural Science Foundation of China (Grant Nos.61631007, 61571117, 61875133, 11874269), the 111 Project (GrantNo.111-2-05) and in part by the China Postdoctoral Science Foundation (Grant No. 2018M633129).

**Authors' Contributions**

S.L., S.J.M. and C.Y. contributed equally to this work. S.L., S.J.M. and W.L.G. carried out the analytical modelling, numerical simulations. S.L., C.Y. and L.Z. completed the sample fabrication and circuit measurements. As the principal investigators of the projects, S.Z., T.J.C., and Y. J. X. conceived the idea, suggested the designs, planned, coordinated and supervised the work. S. L., S. J. M., and S.Z. contributed to the writing of the manuscript. All authors discussed the theoretical and numerical aspects and interpreted the results.

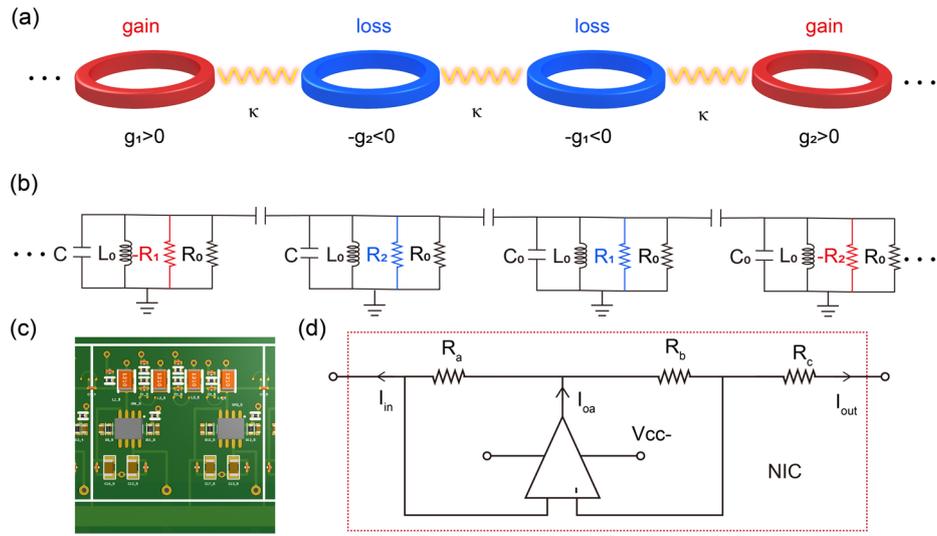

FIG. 1. Schematic and circuit diagram of the gain and loss induced non-Hermitian circuit. **(a)** Schematic of a single cell of the gain and loss induced non-Hermitian model in the quantum system. **(b)** Unit cell of the circuit implementation. **(c)** PCB layout of one unit cell of the non-Hermitian circuit. **(d)** Circuit schematic of the negative resistor module.

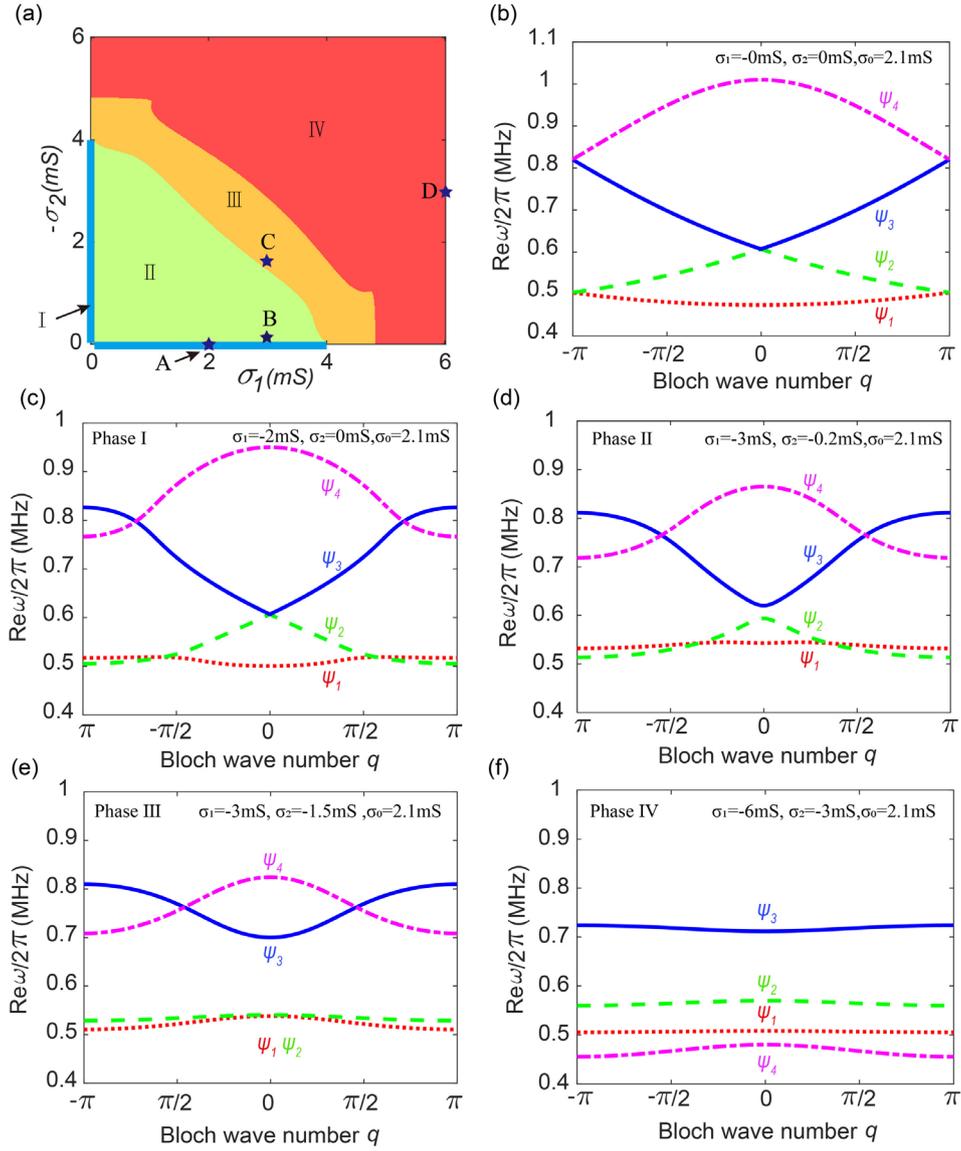

FIG. 2. Phase diagram and the band structures of the four phases for the bulk circuit model. **(a)** Numerically obtained phase diagram of the bulk circuit model with global loss offset $\sigma_0$=2.1mS, which is divided into four different phase regions in the $\sigma_1$-$\sigma_2$ plane ranging from 0 to -6mS. Blue lines located on the *x* and *y* axes represent phase I. **(b)** Bulk band structure Re $\omega(q)$ with $\sigma_1$=0, $\sigma_2$=0, $\sigma_0$=2.1mS. **(c)** Re $\omega(q)$ with $\sigma_1$=-2mS, $\sigma_2$=0mS, $\sigma_0$=2.1mS, phase I. **(d)** Re $\omega(q)$ with $\sigma_1$=-3mS, $\sigma_2$=*-0.2mS*, $\sigma_0$=2.1mS, phase II. **(e)** Re $\omega(q)$ with $\sigma_1$=-3mS, $\sigma_2$=-1.5mS, $\sigma_0$=2.1mS, phase III. **(f)** Re $\omega(q)$ with $\sigma_1$=-6mS, $\sigma_2$=-3mS, $\sigma_0$=2.1mS, phase IV.

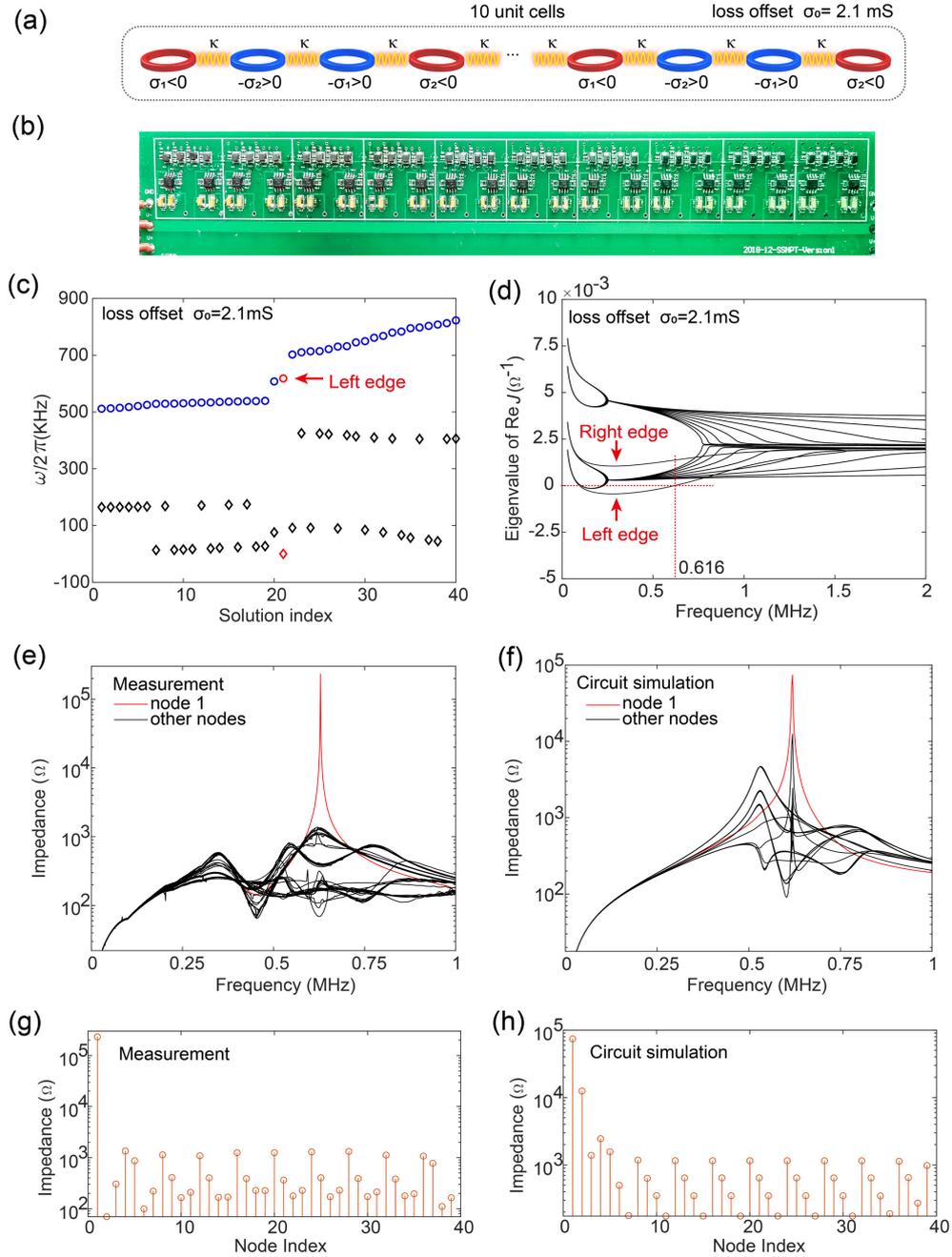

FIG. 3. Experimental and theoretical results for the left edge mode. **(a)** Schematic illustration of the finite circuit chain containing 10 unit cells with $\sigma_1$=-3mS, $\sigma_2$=-1.5mS, $\sigma_0$=2.1mS. **(b)** Fabricated sample. **(c)** Sorted eigenfrequencies of the finite circuit chain. The imaginary part of left edge mode (red) is shifted to zero through the global loss offset $\sigma_0$=2.1mS. Circles and diamond markers represent the real part and imaginary part of the eigenfrequencies, respectively. **(d)** Real part of the eigenvalue spectra of $J(\omega)$ for the finite circuit chain. The two isolated curves represent the two edge modes. The global loss offset $\sigma_0$=2.1mS shifts the lower isolated curve to zero admittance at the midgap frequency (616 KHz). **(e,f)** Experimentally measured and simulated (frequency domain solver) impedance spectra of the finite circuit chain, respectively. Red curve indicates the impedance measured across the leftmost coupling capacitors. **(g,h)** Experimentally measured and simulated (frequency domain solver) impedance distributions of the finite circuit chain.

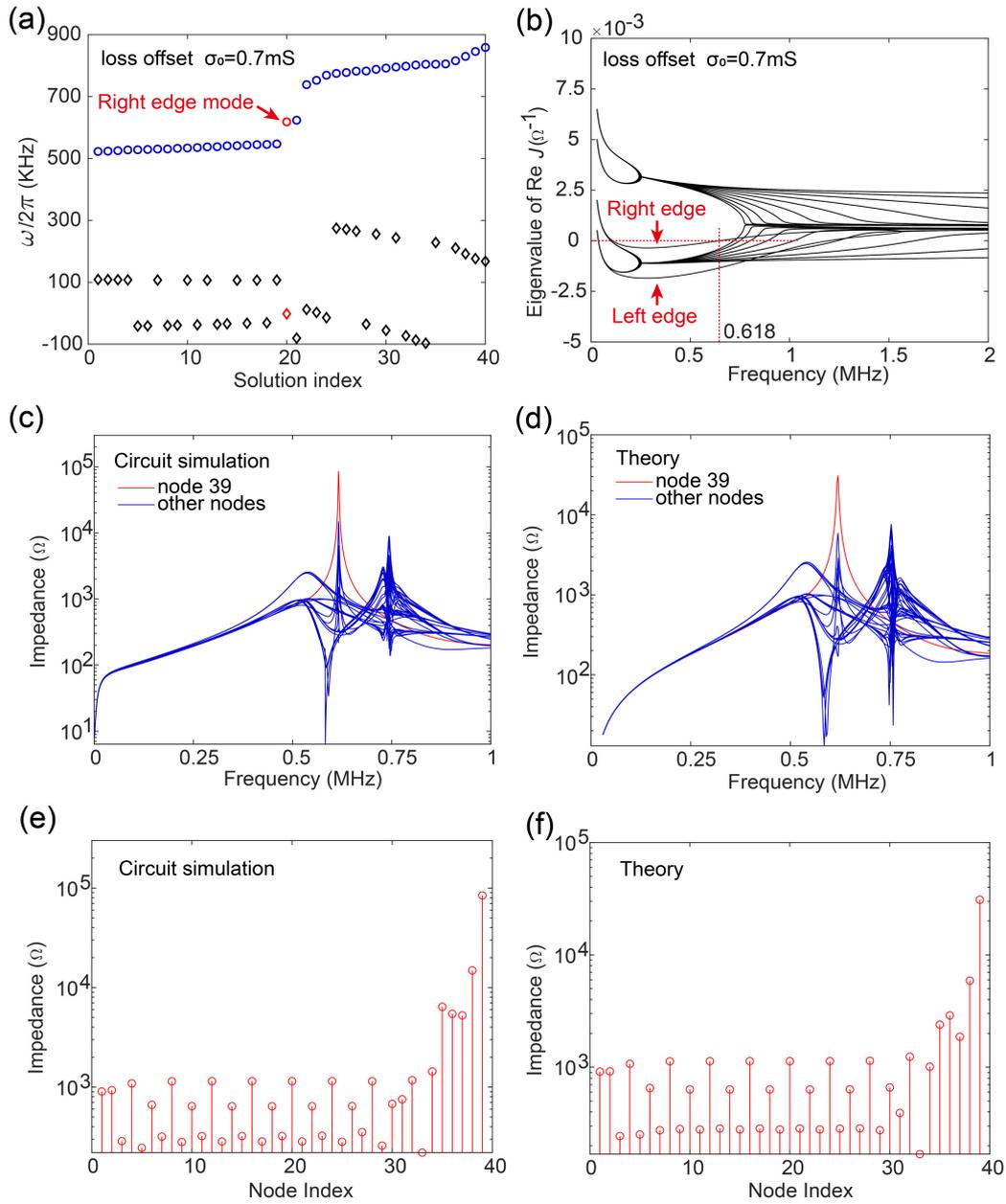

FIG. 4. Numerical and theoretically calculated results for the right edge mode. **(a)** Sorted eigenfrequencies of the finite circuit chain. The imaginary part of left edge mode (red) is shifted to zero through the global loss offset $\sigma_0$=0.7mS. **(b)** Real part of the eigenvalue spectra of $J(\omega)$ for the finite circuit chain. The upper isolated curve representing the left edge mode crosses zero at 618 KHz. **(c,d)** Numerically simulated (frequency domain solver) and theoretically calculated impedance spectra of the finite circuit chain, respectively. Red curve indicates the impedance measured across the leftmost coupling capacitors. **(e,f)** Numerically simulated (frequency domain solver)

and theoretically calculated impedance distributions of the finite circuit chain, respectively.

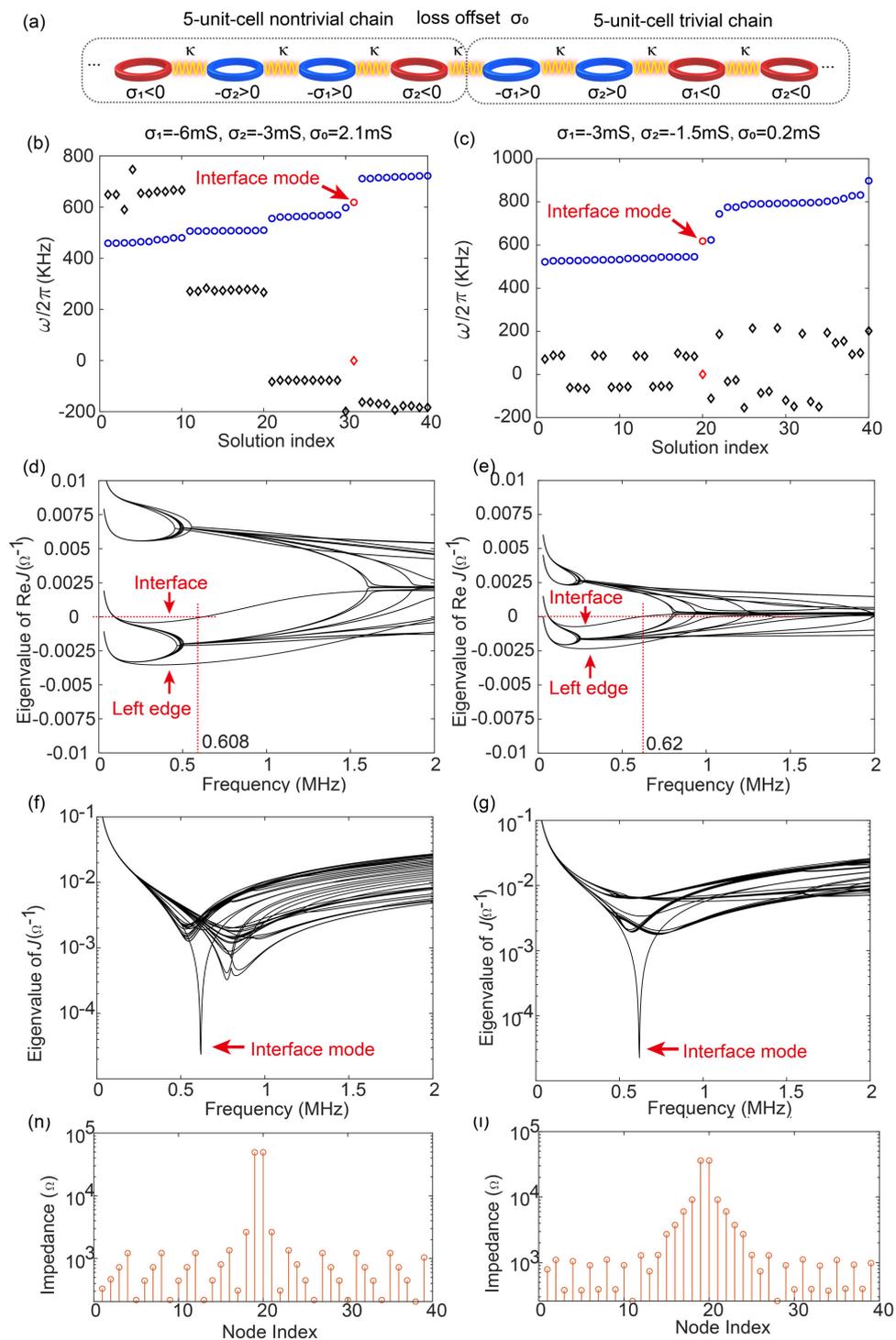

FIG. 5. Demonstration of topological interface mode concentrated at the interface between a trivial and a nontrivial circuit chain. **(a)** Schematic illustration of the finite

circuit chain with a five-unit-cell nontrivial circuit chain and a five unit-cell trivial circuit. The trivial circuit is realized by inverting the sign of $\sigma_1$. **(b,d,f)** Sorted eigenfrequencies, real part of the eigenvalue of circuit Laplacian $J(\omega)$, magnitude of the eigenvalue of circuit Laplacian $J(\omega)$ for the first case with $\sigma_1$=-6mS, $\sigma_2$=-3mS, $\sigma_0$=2.1mS, respectively. **(c,e,g)** Sorted eigenfrequencies, real part of the eigenvalue of circuit Laplacian $J(\omega)$, magnitude of the eigenvalue of circuit Laplacian $J(\omega)$ for the first case with $\sigma_1$=-3mS, $\sigma_2$=-1.5mS, $\sigma_0$=0.2mS, respectively. and theoretically calculated impedance distributions, respectively. **(h,i)** Theoretically calculated impedance distributions for the first and second cases, respectively.